\documentclass[aps,prb,twocolumn,preprintnumbers,amsmath,amssymb,superscriptaddress
]{revtex4-1}
\usepackage{graphicx}
\usepackage{subfigure}
\usepackage{epsfig}
\usepackage{dcolumn}
\usepackage{bm}
\usepackage{ulem}
\usepackage{color}
\usepackage{hhline}

\def\be{\begin{equation}}       \def\ee{\end{equation}}
\def\bea{\begin{eqnarray}}      \def\eea{\end{eqnarray}}
\def\ba{\begin{array}}
\def\ea{\end{array}}
\def\bnum{\begin{enumerate} }
\def\enum{\end{enumerate}}

\def\nn{\nonumber}

\def\=>{\Rightarrow}
\def\>{\rightarrow}
\def\A{\uparrow}
\def\V{\downarrow}

\def\eye2{Fathbb{I}}

\def\Eq#1{Eq.~(\ref{#1})}

\def\Tr{\mathrm{Tr}}

\renewcommand{\>}{\rangle}

\begin{document}

\title{Fermion-induced quantum critical points in three-dimensional Weyl semimetals}

\author{Shao-Kai Jian}
\affiliation{Institute for Advanced Study, Tsinghua University, Beijing 100084, China}

\author{Hong Yao}
\email{yaohong@tsinghua.edu.cn}
\affiliation{Institute for Advanced Study, Tsinghua University, Beijing 100084, China}
\affiliation{State Key Laboratory of Low Dimensional Quantum Physics, Tsinghua University, Beijing 100084, China}
\affiliation{Collaborative Innovation Center of Quantum Matter, Beijing 100084, China}

\begin{abstract}
Fermion-induced quantum critical points (FIQCPs) were recently discovered at the putatively first-order transitions between two-dimensional (2D) Dirac semimetals and the Kekule valence bond solids on the honeycomb lattice by {\it sign-free} quantum Monte Carlo simulations [Nature Communications {\bf8}, 314, (2017)]. Here,  we investigate possible FIQCP in 3D topological Weyl semimetals at a $Z_3$ symmetry-breaking transition that is putatively first-order according to the Landau criterion. We construct a lattice model featuring 3D double-Weyl fermions (monopole charges $\pm$2) and we show that $Z_3$ {\it nodal-nematic} transitions occur under finite Hubbard interaction. Furthermore, using renormalization-group analysis, we identify such a transition as a genuine FIQCP where the cubic terms are irrelevant and an enlarged U(1) symmetry emerges at low energy. We further discuss quantum critical behaviors and experimental signatures of such FIQCPs in 3D double-Weyl semimetals.
\end{abstract}
\date{\today}

\maketitle

\section{Introduction}		

The nature of a quantum phase transition is strongly dictated by the symmetry of the order parameters and the spatial dimensions of the systems in question \cite{sachdevbook}. One textbook criterion according to Landau \cite{landau1958,lifshitz1942} states that if cubic terms of order parameters form a trivial representation of the symmetry group of the systems, the phase transition is necessarily first-order. This is most easily seen from the fact that the order parameter will develop a finite jump through the phase transition, if the Landau-Ginzburg (LG) free energy includes cubic terms of order parameters. Previous work showed that this mean-field criterion works well in three dimensions or higher \cite{blote1979, wu1982, binder1987}.

One may wonder whether and where phase transitions that violate the cubic-term criterion discussed above can occur, since deconfined quantum critical points (DQCPs)  \cite{senthil2004a, sandvik2007, melko2008, kaul2013, lou2009, pujari2013, nahum2015, senthil2004b, wang2015,shao2016} have provided a novel way of realizing quantum phase transitions that violate the Landau criterion of first-order transitions between two symmetry-incompatible phases. One intriguing scenario violating the cubic-term criterion was provided by strong fluctuations in low dimensions: the quantum three-state Potts model in 1+1D (equivalently the classical three-state Potts model in two dimensions) is an exactly solvable model being a well-known example that violates Landau's cubic-term criterion \cite{baxter1973}. Recently, a distinct and higher-dimensional scenario was introduced: quantum phase transitions in fermionic systems \cite{li2015a}.

At zero temperature, gapless fermionic degrees of freedom must be retained in quantum LG theory, and their presence at quantum phase transitions may dramatically change the nature of critical behaviors. Although modifications of critical behaviors by gapless fermions have been studied extensively \cite{hertz1976, moriya1985, millis1993, fradkin2001, chubukov2003, lohneysen2007, max2010a, max2010b, berg2016, lee2016, rosenstein1993, herbut2006, kivelson2008, sachdev2008}, it was shown only recently in Ref. \cite{li2015a} by both large-scale Majorana quantum Monte Carlo (QMC) \cite{li2015b,li2016} simulations and large-$N$ renormalization group (RG) analysis that gapless Dirac fermions can drive a putatively first-order quantum phase transition between two-dimensional (2D) Dirac semimetals and the Kekule valence bond solids (Kekule-VBS) into a continuous one, which is called a fermion-induced quantum critical point (FIQCP). Such a FIQCP was also confirmed by a more recent RG analysis using $\varepsilon$-expansion \cite{scherer2016}.

\begin{figure}
	\includegraphics[width=7.2cm]{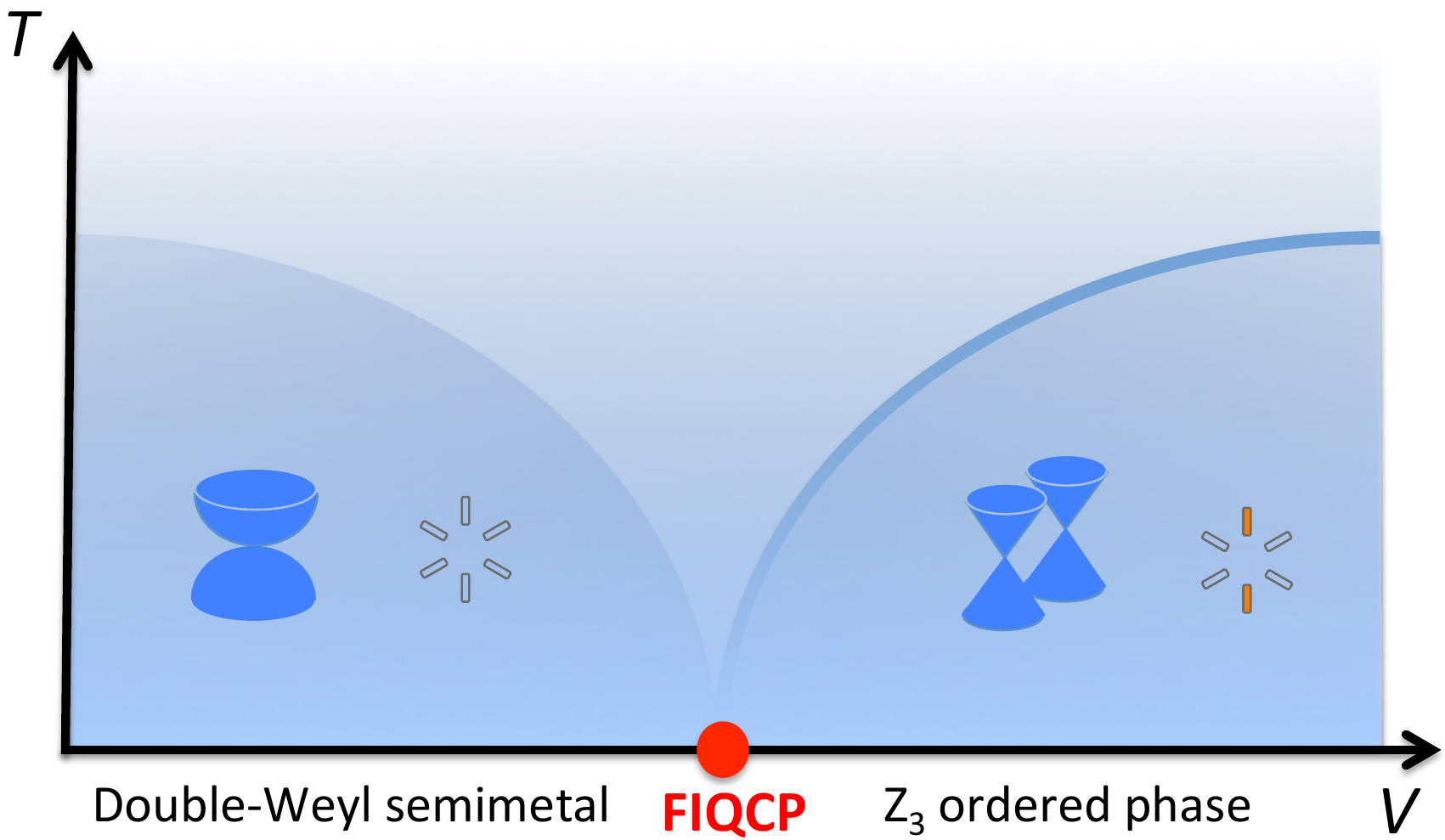}
	\caption{\label{fig1}A schematic phase diagram for a $Z_3$ nodal-nematic transition from a topological double-Weyl semimetal to a nematic phase where each double-Weyl point splits into two Weyl points. A FIQCP emerges at zero temperature while the transition at finite temperature is still first-order.}
\end{figure}

When symmetry-breaking happens in a system with gapless fermions, they experience different fates. For instance, the Kekule-VBS order breaks translation symmetry, and gaps out Dirac fermions in the ordered phase \cite{roy2013,li2015a,jian2015a,wu2016,lang2013,motruk2015,wu2016b}. A nodal-nematic order, on the other hand, does not gap out nodal fermions but shifts the positions of the nodes in $k$-space \cite{kivelson2008, sachdev2008}. Here, we investigate if FIQCP can occur at a $Z_3$ nodal-nematic phase transition in 3D topological double-Weyl semimetal \cite{xu2011, fang2012, jian2015b, lai2015, dai2015, huang2016, moon2013, nagaosa2014a, sun2009}, where a $Z_3$ order parameter cannot gap out the fermions due to non-vanishing monopole charge ($\pm$2) of double-Weyl points. Instead, when nematic orders form, each double-Weyl point splits into two Weyl points with monopole charge $\pm$1 \cite{ashvin2011,burkov2011,weng2015,hasan2015,xu2015,lv2015,yang2015}, partially breaking the rotational symmetry $C_6$ to $C_2$. At such a transition, cubic terms of the order-parameter are allowed in quantum LG free energy; nonetheless, we show that the putative first-order phase transition can be driven into a continuous one, i.e., a FIQCP. A schematic phase diagram for the occurrence of such a FIQCP is shown in Fig. \ref{fig1}.

\section{Lattice model}

We first consider an interacting microscopic model of double-Weyl fermions featuring $Z_3$ nodal-nematic phase transitions. Specifically, we construct an interacting spin-1/2 electron model on a 3D hexagonal lattice with lattice vectors $\vec a_1$=$(1,0,0)$, $\vec a_2$=$(-\frac12, \frac{\sqrt{3}}{2},0)$, and $\vec a_3=(0,0,1)$, where the lattice constants both in the triangle plane and along the $c$-axis are set to unity. The Hamiltonian is given by
\bea
H= \sum_{\vec k} c_{\vec k}^\dag [d_x\sigma^x+d_y\sigma^y+d_z\sigma^z] c_{\vec k} + U \sum_{i} c_{i\A}^\dag c_{i\A} c_{i\V}^\dag c_{i\V},~~~~
\eea
where $c^\dag_{\vec k}$=$(c^\dag_{\vec k\A},c^\dag_{\vec k\V})$ are creation operators of spin-1/2 electrons in momentum space, $d_x(\vec k)$=$-2t_1(\cos k_1-\frac12 \cos k_2- \frac12 \cos k_3)$, $d_y(\vec k)$=$-2t_1( \frac{\sqrt{3}}{2} \cos k_2-  \frac{\sqrt{3}}{2} \cos k_3)$, $d_z(\vec k)$=$ -2t_2 (\cos k_1+ \cos k_2 + \cos k_3)-2t_3 \cos k_z +m$. Here $k_i=\vec k \cdot \vec a_i$, $t_j$ are hopping amplitudes, $m$ is a Zeeman coupling, and $\sigma^j$ are Pauli matrices with spin indices. $U$ is the strength of the onsite Hubbard interactions.

It is clear that the Hamiltonian on the three-dimensional hexagonal lattice above is invariant under the $C_6$ rotation along the $z$-axis. When $6t_2-2 t_3<m<6t_2+ 2t_3$, there are two double-Weyl points located at $(0,0, \pm K )$ with $K\!=\! \arccos \frac{m-6t_2}{2t_3}$. The double-Weyl points at the non-interacting limit are protected by the $C_6$ symmetry of the hexagonal lattice. Owing to inversion symmetry, these two double-Weyl points are located at the same energy. For the non-interacting part $H_0=\sum_{\vec k} c^\dag_{\vec k} h(\vec k) c_{\vec k}$, one obtains the following low-energy continuum description by expanding $h(\vec k)$ around two double-Weyl points:
\bea
	h_0(\vec k)= A[(k_x^2-k_y^2)\sigma^x+2k_x k_y \sigma^y]+v_{f3} k_z \sigma^z \tau^z,
\eea
where $A= \frac{3}{4} t_1, v_{f3}=\sin K$ and $\tau$ are Pauli matrices acting on the valley basis.

The double-Weyl points are robust against weak interaction $U$. However, when the repulsive $U$ is sufficiently strong, the system could be unstable towards nematic order, which breaks the $C_6$ symmetry down to $C_2$ and causes each double-Weyl point to split into two Weyl points.  This can be heuristically understood as the density of states around Weyl points, $\rho(\epsilon) \propto \epsilon^2 $, is smaller than that around double-Weyl points, $\rho(\epsilon) \propto \epsilon$. The $Z_3$ nematic order, $\phi_i(x) \!=\!\langle c^\dag(x) \sigma^i c(x) \rangle, i \!=\!1,2$, is a doublet (two-component real boson) in the $E_2$ representation.

\begin{figure}
\subfigure[]{\label{insulator_order}
\includegraphics[height=2.5cm]{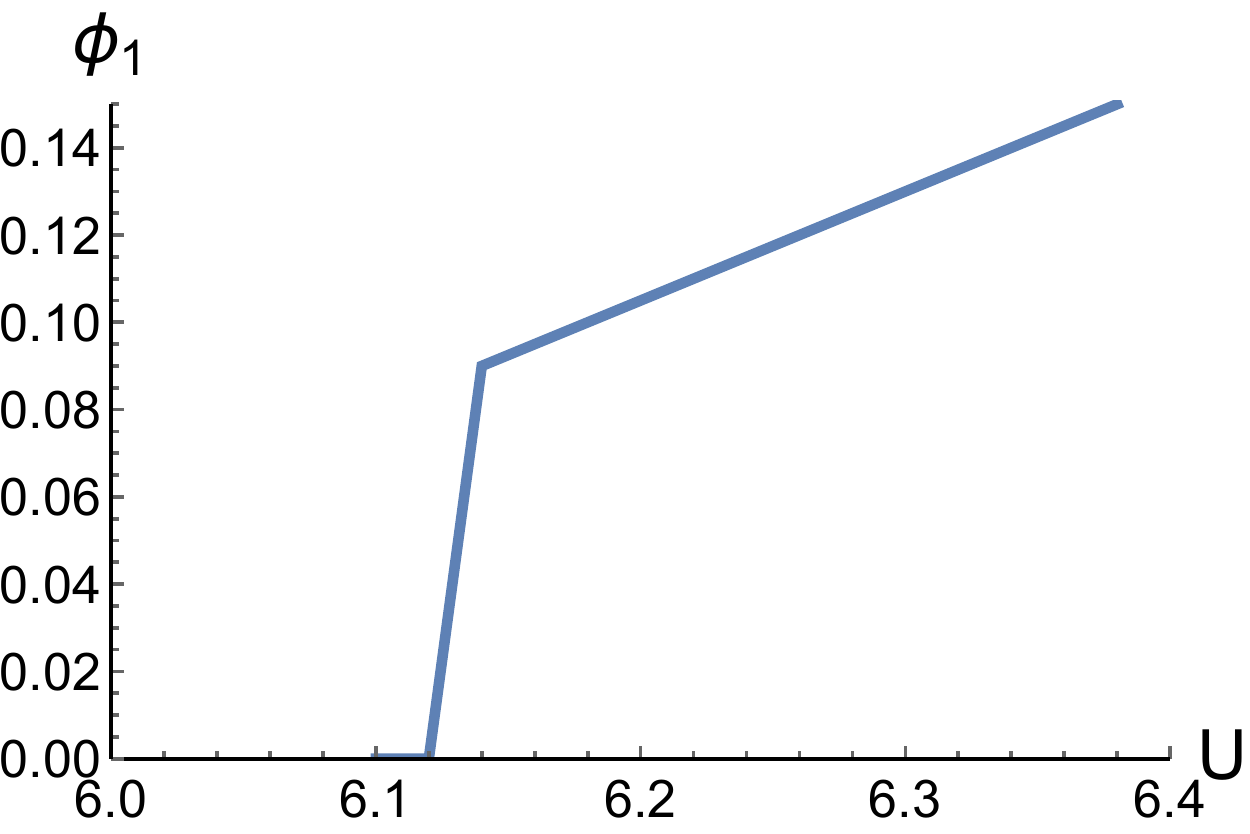}}
\subfigure[]{\label{dwf_order}
\includegraphics[height=2.5cm]{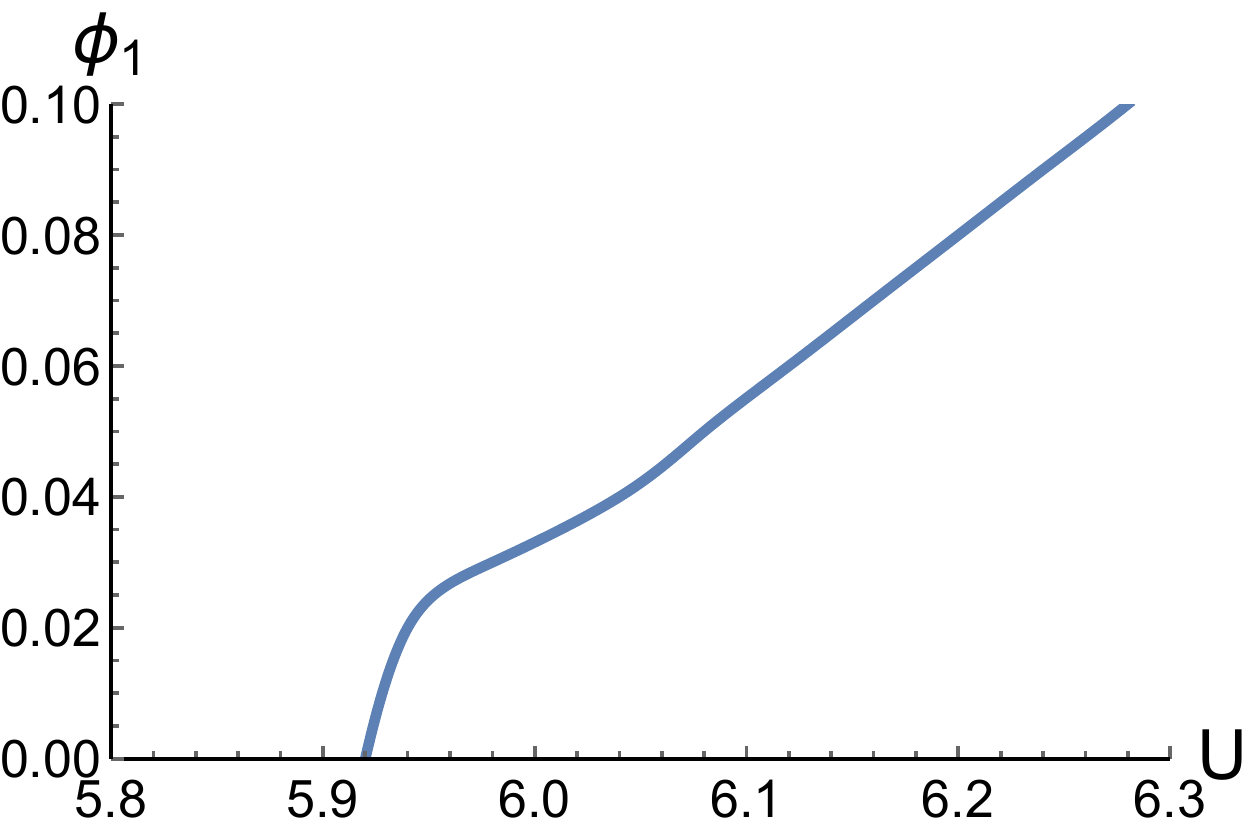}}
\caption{The $Z_3$ nematic order is analyzed through mean-field calculations setting $t_1\!=\!t_2\!=\!t_3\!=\!1$. (a) The order parameter as a function of $U$ for the case of $m=3.9$ whose dispersions is fully gapped describing an insulator. The $Z_3$ quantum phase transition is clearly a first-order. Part (b) shows that in a gapless system, $m=4.1$.}
\end{figure}

To see if the $Z_3$ nematic order occurs, we perform mean field calculations of the phase diagram as a function of $U$, using the parameters $t_1\!=\!t_2\!=\!t_3\!=\!1$ (see the Appendix A for details). For comparison, we study two cases: $m\!=\!3.9$ and $m\!=\!4.1$. Note that for the former choice of $m$, the spectra of $h(\vec k)$ are actually gapped and the system is an insulator, while for the latter there are two double-Weyl points locating on the $k_z$ axis. The $Z_3$ order parameter as a function of $U$ is shown in \ref{insulator_order} and Fig. \ref{dwf_order}, respectively. For the insulator case ($m\!=\!3.9$) where there is no gapless fermion affecting the qualitative behaviors of the phase transitions, it is clear that there is a finite jump in the order parameter around $U\approx 6.14$, clearly indicating a first-order transition, which is expected according to the cubic-term Landau criterion.  On the other hand, for the double-Weyl fermion case ($m\!=\!4.1$) where gapless fermions may qualitatively alter the nature of the phase transitions, the $Z_3$ nematic order also appears when $U \!>\! 6.10$; however, it looks dramatically different from the first-order behavior in \ref{insulator_order}. Within numerical accuracy, it looks like a continuous transition from the simple mean field analysis, indicating that the presence of gapless fermions reduces the signature of first-order transition. Note that the mean-field analysis of the nature of the phase transition may not capture the nature of phase transitions at strong $U$. For such a $Z_3$ nodal-nematic transition, since the low-energy physics involves gapless fermions, one should treat quantum fluctuations of fermions and bosons on an equal footing via RG calculations.

\section{Effective theory}

The effective Lagrangian near the $Z_3$ nodal-nematic transition point consists of gapless double-Weyl fermions $\psi$, a fluctuating $Z_3$ order parameter $\phi$,
and the coupling terms between them, i.e., $\mathcal{L} \!=\! \mathcal{L}_\psi+\mathcal{L}_\phi+\mathcal{L}_{\psi\phi}$. The double-Weyl fermion action is given by
\bea
\mathcal{L}_\psi \!=\! \psi^\dag  [- i\omega \!+\! h_0(\vec k)] \psi,
\eea
where $\psi=(\psi_+, \psi_-)^T$, $\psi_\pm$ are two-component double-Weyl fermions at the $\pm K$ valley, respectively, and $\omega$ is the Matsubara frequency. The dispersion of double-Weyl fermions is anisotropic and gives rise to monopole charge $\pm 2$ in $k$-space. The $Z_3$ nematic order, ($\phi_1$,$\phi_2$) can be described by a complex boson $\phi \!\equiv\! \phi_1\!-\! i\phi_2$. A $C_6$ rotation simply takes $\phi$ to $e^{-i\frac{2\pi}{3}} \phi$ such that its cubic term is allowed in the effective theory. The effective Lagrangian for the $Z_3$ order-parameter fields is given by
\bea\label{bosonaction}
	\mathcal{L}_\phi\! &=& \! |\partial_\tau \phi|^2 \!+\! v_{b\perp} ^2 \sum_{i=1}^2|\partial_i \phi|^2 \! +\!v_{b3}^2 |\partial_z \phi|^2 \nn \\
	&&~~~~~~~~~~~~~~+ r |\phi|^2 \!+\! b(\phi^3\!+\!\phi^{\ast3}) \!+\! u|\phi|^4,
\eea
where $v_{b\perp}$ and $v_{b3}$ denote the boson velocity in the $xy$-plane and along the $z$-axis, respectively. $r$ is a boson mass that tunes the phase transition, and $b, u$ are the strength of the cubic and quartic terms, respectively. Non-vanishing $b$ is allowed in \Eq{bosonaction}, putatively rendering a first-order transition according to the Landau criterion. (Note that a $Z_3$ phase transition out of a topologically-ordered phase \cite{cenke2011} can be driven by condensing fractionalized anyons whose LG theory is qualitatively different from \Eq{bosonaction}.)  Moreover, the double-Weyl fermions and the order parameter fluctuations are coupled. The effective coupling is dictated by symmetries, and it is given by
\bea
	\mathcal{L}_{\psi\phi}= g(\phi \psi^\dag \sigma^+  \psi + \phi^\ast \psi^\dag \sigma^- \psi),
\eea
where $g$ is a real Yukawa coupling constant and $\sigma^\pm \!=\! \frac{1}{2} (\sigma^x \!\pm\! i \sigma^y)$.

Owing to the non-vanishing monopole charge of a double-Weyl point, breaking rotational symmetry does not gap out the fermions. For instance, assuming $\langle\phi\rangle \!=\! \frac{m}{gA}\!>\!0$ in the ordered phase, the dispersion of fermions is then given by
\bea
E_k\!=\! \pm \sqrt{A^2[ (k_x^2-k_y^2+ m)^2 +4k_x^2 k_y^2]+ v_{f3}^2 k_z^2} ,
\eea
from which one can deduce that the double-Weyl point at $(0,0,K)$ is split into two Weyl points located at $\vec k= (0,\pm\sqrt{m},K)$ and similarly for the other double-Weyl point at $(0,0,-K)$.

\section{Renormalization group analysis}

We now present strong evidence of a FIQCP at the $Z_3$ nodal-nematic transition in double-Weyl semimetals by performing standard RG analysis in which fermions and bosons are treated on an equal footing. The RG procedure is to integrate out fast modes to generate RG equations \cite{sachdevbook, moshe2003}. In calculating the RG equation, we generalize the fermion to $N$ flavors ($N$ denotes the number of the four-component double-Weyl fermions).

A subtlety arises due to anisotropic dispersion of double-Weyl fermions, i.e., the scaling properties of orthogonal spatial directions are different \cite{jian2015b, lai2015, nagaosa2014b, nagaosa2016}. Here, we assume the scaling dimension for the three momenta and the frequency to be $[k_z]\!=\!1,[k_{x,y}]\!=\!z_1,[\omega]\!=\!z$ without loss of generality, where $[\cdots]$ denotes the scaling dimension. The values of $z$, $z_1$ as well as anomalous dimensions are determined by renormalization of the kinetic part of the action, i.e.,
\bea
	\delta S^{(1)}=\int \frac{d^4p}{(2\pi)^4} \Big[ \psi^\dag(p) \Sigma(p) \psi(p)+ \phi^\ast(p) \Pi(p) \phi(p) \Big], \label{kine}
\eea
where $\Sigma$ and $\Pi$ are fermion and boson self-energies resulting from integrating out the fast modes in the momentum shells (the momentum shell is chosen to be an "infinite cylinder" with radius $\Lambda_\perp$, see Appendix C for details).

From the Eq.(\ref{kine}), we can obtain the RG equations of velocities. To simplify the analysis, we assume the velocity difference between $v_{b3}$ and $v_{f3}$ is small, and we let $\frac{v_{b3}}{v_{f3}}\!=\! 1\!+\! \lambda$, with $|\lambda|\! \ll \!1$. The RG equations for $\lambda$ is given by $\frac{d \lambda}{dl}= -\Delta_\lambda \lambda \label{delta}$, where $l\!>\!0$ is the flow parameter, and $\Delta_\lambda$ is a positive constant independent of $\lambda$ (see Appendix C for details). As a result, $\lambda=0$ is a stable fixed point. In the other words, the boson and fermion velocity along $z$-axis, $v_{b3}$ and $v_{f3}$, will flow to the same value in low energy for small $\lambda$. Thus in the following we set $v_{f3}\!=\!v_{b3}\!=\!v$ for simplicity. And the RG flow of velocity $v$ is controlled by the dynamical critical exponent $z$, i.e., $\frac{d \log v}{d l} \!=\! z\!-\!1$. Since the velocities are physical observables, they must stay finite and this requires $z\!=\!1$ at the fixed point.

For later simplicity in expressing the RG equations, we introduce four dimensionless coupling constants (not to be confused with critical exponents):
\bea
\beta\!=\!\frac{b^2}{\pi^2 v_{b\perp}^4 v \Lambda_\perp^2},~~\gamma \!=\!\frac{g^2}{24\pi^2 A^2 v \Lambda_\perp^2},~~ \delta\!=\! \frac{u}{\pi^2 v_{b\perp}^2 v},
\eea
corresponding to three running coupling constants, i.e., $b$, $g$ and $u$ in the interacting Lagrangian, respectively and $\alpha\!=\!\frac{A\Lambda_\perp}{v_{b\perp}}$. Then the RG equation for boson velocity in the $xy$-plane $v_{b\perp}$ reads (see Appendix C for details),
\bea
	\frac{d\log v_{b\perp}}{dl} &=& 1-z_1-\frac38 \beta+N (2\alpha^2-1) \gamma. \label{vbp}
\eea
In a similar way, $z_1$ can also be determined at the fixed point from the RG equation for boson velocity in $xy$-plane  $z_1\!=\!1\!-\!\frac{3}{8}\beta+N (2\alpha^2-1) \gamma$. Moreover, the anomalous dimensions for fermions and bosons are also obtained from Eq.(\ref{kine}): $\eta_\psi\!=\! \frac{3(1\!-\!\alpha^2\!+\!\alpha^2 \!\log \alpha^2)}{2(1-\alpha^2)^2} \alpha^2 \gamma$ and $\eta_\phi \!=\! N\gamma \!+\! \frac{3}{16}\beta$.

\begin{figure}
	\includegraphics[width=5.cm]{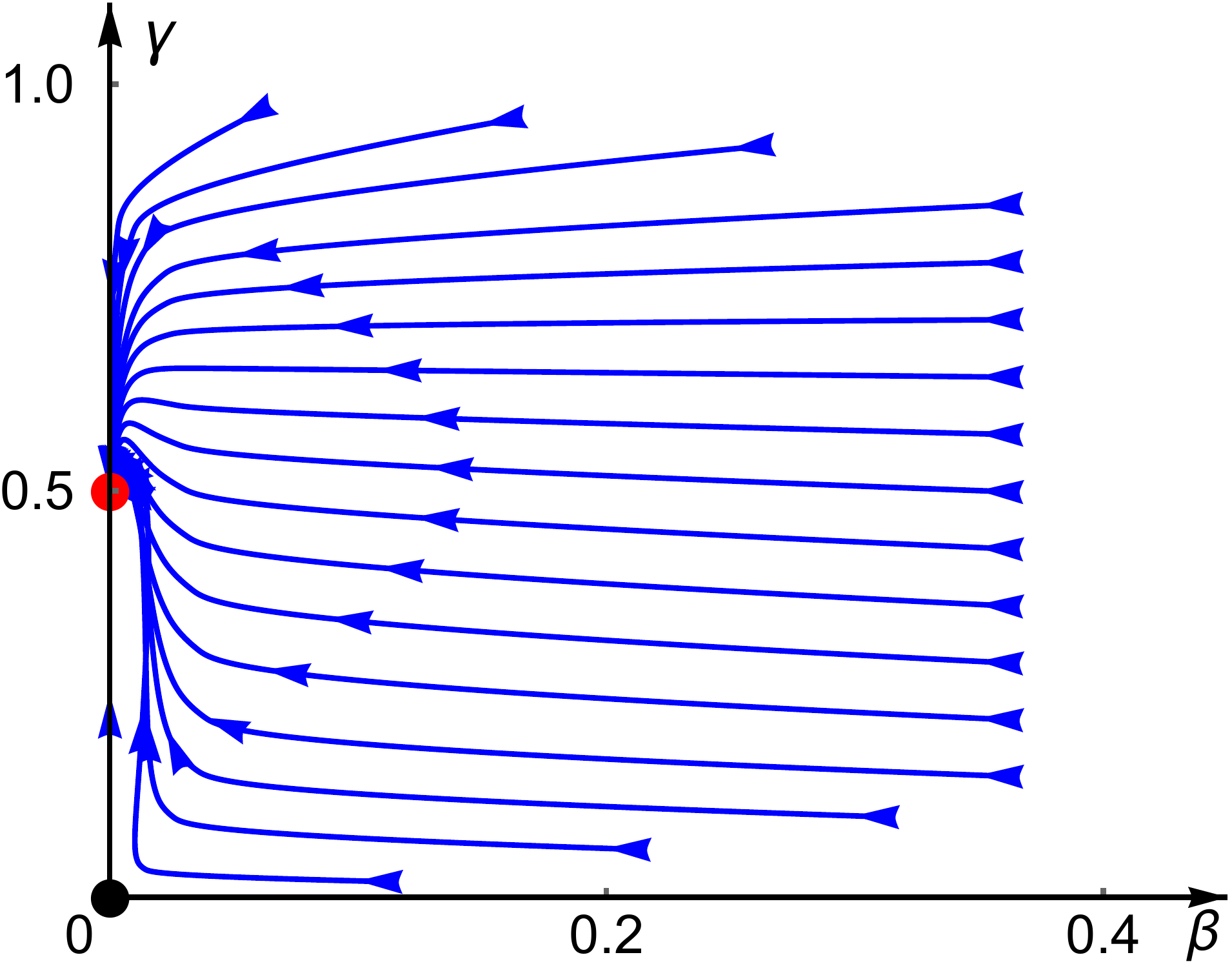}
	\caption{\label{fig6} The flow diagram $\beta$-$\gamma$ for the $Z_3$ transition in double-Weyl semimetals for $N=1$ case. The arrowed curves indicate the running coupling constants as a function of energy. The red and black circles located at (0,$\frac12$) and (0,0) indicate a stable fixed point and a Gaussian fixed point, respectively. The red one is identified as a fermion-induced quantum critical point. At these fixed point, one gets $\alpha^*=\delta^*=0$.}
\end{figure}

After getting the expressions of dynamical exponents $z$'s and anomalous dimensions $\eta$'s near a physical fixed point, we are now in a position to analyze the RG equations of dimensionless coupling constants resulting from renormalization of the interaction part in the action, i.e.,
\bea
	\delta S^{(2)}= \int d^4 x \Big[ \Gamma_{\phi^3} \phi^3+  \Gamma_{\phi^{\ast3}} \phi^{\ast3}+ \Gamma_{|\phi|^4} |\phi|^4 \Big].
\eea
These vertices $\Gamma$'s are evaluated in Appendix C. Note that the vertex between the fermion and the boson is not renormalized. After obtaining the RG equations for various coupling constants, such as $g$, $b$ and $u$, we convert the RG equations to that of dimensionless coupling constants.

We state the main results here; those readers who want to know the full RG equations, please should refer to Appendix C for details. There are two fixed points with both $\gamma\!\ge\!0$ and $\delta\!\ge\!0$: one is the usual Gaussian fixed point and the other is a nontrivial fixed point given by $(\alpha^*,\beta^*,\gamma^*,\delta^*)$=(0,0,$\frac{1}{2N}$,0), as shown in Fig. \ref{fig6}, where the RG flows in $(\beta,\gamma)$-plane are drawn. $N$ is the number of four-component double-Weyl fermions, and $N\!=\!1$ corresponds to the lattice system we introduced before. As indicated in Fig. \ref{fig6}, the Gaussian fixed point at the origin is unstable, while the fixed point at $(\beta^*, \gamma^*)$=(0,$\frac{1}{2N}$) is stable.

Note that $\alpha$ approximately captures the ratio of kinetic energy between fermions and bosons in $xy$-plane. When the system approaches the nematic transition from the disordered phase, fermion dispersion along the splitting direction becomes soft, and one can approximate the RG equations to the lowest order of $\alpha$ \cite{kivelson2008,sachdev2008}. To further justify this, the RG equations near $\alpha=0$ at the stable fixed point read
\bea
	\frac{d \alpha}{dl} =-(2+ \frac{3}{2N}) \alpha^3,
\eea
which shows that $\alpha$ is irrelevant at this fixed point. Under this approximation, one gets simplified RG equations to the lowest order of $\alpha$ near the fixed point,
\bea
	\frac{d \beta}{dl} &=& (2-4N\gamma)\beta-\frac38 \beta^2, \\
	\frac{d \gamma}{dl} &=& (2-4N\gamma)\gamma, \\
	\frac{d \delta}{dl} &=& -(2N\gamma+\frac54 \delta)\delta.
\eea
Apparently, a Gaussian fixed point is one solution of the RG equations shown above. However, it is unstable against the perturbations along the $\beta$ and $\gamma$ directions. There is a stable fixed point, as already indicated in the flow diagram in Fig. \ref{fig6}, at $(\gamma^*\!,\beta^*\!,\delta^*)$=($\frac{1}{2N}$,0,0). The eigenvalues of the stability matrix are $(0,-2,-1)$, where the zero eigenvalue indicates a marginal direction. Indeed, one finds that the deviation $\Delta \beta$ along the $\beta$ direction is marginally irrelevant, i.e., $\frac{d\Delta \beta}{dl} = -\frac38 (\Delta\beta)^2$. Note that $\beta \ge 0$ by definition. Thus, the nontrivial fixed point is irrelevant under perturbations along the $\gamma$ and $\alpha$ directions and marginally irrelevant under perturbations along $\beta$ direction. A stable fixed point at the critical surface corresponds to a genuine continuous critical point. At this nontrivial stable fixed point, one finds that $b^2\! \propto\! \beta\!=\! 0$, i.e. the cubic terms of the $Z_3$ order-parameter are irrelevant. Consequently, this fixed point corresponds to a continuous phase transition, namely, a FIQCP! Moreover, the system has an emergent $U(1)$ symmetry (the rotation of the system along $z$-axis) at the FIQCP.

The anomalous dimensions for fermions and bosons at the nontrivial fixed point are given by $\eta_\psi \!=\! 0$ and $\eta_\phi \!=\! \frac12$ yielding the critical exponent $\eta\!=\!2\eta_\phi \!=\!1$. Though the FIQCP is distinguished with the Gaussian fixed point, the vanishing fermion anomalous dimension implies that the quasiparticle picture is still valid, in contrast to the FIQCP in two-dimensional Dirac fermions \cite{li2015a}. Due to the validation of the quasiparticle, one expects that the critical exponent $\nu$ is given by the naive scaling argument $\nu^{-1}=2+2z_1-2[\phi]=1$, where $z_1\!=\!1\!-\!\frac{3}{8}\beta^*\!+\!N (2\alpha^{*2}-1) \gamma^*\!=\!\frac12$. and $[\phi]\!=\!\frac12+\eta_\phi\!=\!1$ is the scaling dimension of a boson field at this fixed point.

We would like to emphasize that it is the presence of gapless fermions that dramatically changes the nature of the $Z_3$ nematic phase transition. If we naively turn off fermions, i.e., set $N=0$, then $\gamma$ disappears from the RG equations of $\beta$ and $\gamma$. Now the fixed point with $\beta=0$ is strongly relevant along $\beta$, which would render a first-order transition, as expected from the Landau criterion. Consequently, we expect that there should exist a critical value $N_c$ such that a FIQCP occurs for $N\!>\!N_c$ and a first-order transition for $N\!<\!N_c$. The current one-loop RG calculations shows that FIQCP occurs for any finite value of $N$, and a more accurate value of $N_c$ may be obtained from higher loop RG analysis in the future.

There is a heuristic argument for the occurrence of such FIQCP at large $N$. Integrating out gapless fermions can result in a non-analytical term, e.g., $|\phi|^3$, of the order parameter, and this term may overwhelm the original cubic terms at the phase transition and drives the first-order transition into a continuous one. To show this explicitly, we implement a simplified method by integrating out fermions all at once and then expanding the effective free-energy as a function of the order-parameter. We find that the effective free energy includes a non-analytical term \cite{imada2013} (see Appendix B for details): $F_{\text{non}}[\phi]= N b'|\phi|^3$, where $N$ is the flavor of four-component double-Weyl fermions and $b'$ is a positive constant depending on the momentum cutoff. If $N \!>\!\frac{2|b|}{b'}$, the cubic term in the free energy has a bound $Nb'|\phi|^3\!+\! b(\phi^3 \!+\! \phi^{*3}) \!\ge\! (Nb' \!+\! 2b) |\phi|^3$, and the minimal energy is achieved  from $\phi\!=\!0$ to nonzero continuously through phase transition. A continuous phase transition can occur at a putative first-order transition as long as the flavors of fermions $N$ is sufficiently large, consistent with RG calculations. Note that the mean-field analysis predicted a wrong critical exponent $\nu=1/2$, which is quite different from the RG result of $\nu=1$ because the former cannot fully capture quantum fluctuations.

\section{Conclusions and discussions}

It is worth pointing out again that the fluctuations of fermions at zero temperature play an essential role in a FIQCP. The three-state Potts model is a neat example featuring a $Z_3$ transition without gapless fermion modes, where the transition is shown to be first-order in 2+1D and higher dimensions \cite{herrmann1979, wu1982}. However, if the transition involves large enough gapless fermions, a FIQCP may occur. The scaling dimension of the order-parameter field is often enhanced by fermions. Indeed, $[\phi]\!=\!1$ at the stable fixed point corresponding to the $Z_3$ nodal-nematic transition in double-Weyl semimetals is larger than the nominal scaling dimension of the order-parameter assigned for the first-order $Z_3$ transitions \cite{fisher1982}. Moreover, it is consistent with the rigorous lower bound of scaling dimension of order-parameter fields, $[\phi]\!>\!0.565$, required to induce an emergent $U(1)$ symmetry from the $Z_3$ symmetry based on recent conformal bootstrap calculations \cite{nakayama2016}. Large anomalous dimension is also a typical feature of DQCP \cite{senthil2004a,senthil2004b,motrunich2004}, where the deconfined spinons play a similar role to that of gapless electrons here.

In conclusion, we construct a 3D lattice model hosting topological double-Weyl semimetal. By tuning onsite Hubbard interactions, the system undergoes a quantum phase transition into a $Z_3$ nodal nematic phase. The phase transition is analyzed through a mean-field calculations: it is first-order without gapless fermions in the system, while weakly first-order or continuous with the presence of gapless fermions. To distinguish the nature of the transition in the presence of gapless fermions, we further present a RG study of the $Z_3$ nodal-nematic transition, where the low-energy effective field theory contains cubic terms of order-parameters. A marginal stable non-trivial fixed point is identified as a FIQCP, at which a marginal Fermi liquid theory is expected. This novel FIQCP may be observed in the future in candidate double-Weyl materials such as the one synthesized by stacking Chern insulators \cite{hughes2016}, and it could lead to a united understanding of quantum critical phenomena.

\section{Acknowledgments}
We would like to thank Xin Dai and Shi-Xin Zhang for helpful discussions. This work was supported in part by the Ministry of Science and Technology of China under Grant No. 2016YFA0301001 (H.Y.) and by the National Natural Science Foundation of China under Grant No. 11474175 (S.-K.J. and H.Y.).

\begin{appendix}

\begin{widetext}

\renewcommand{\theequation}{S\arabic{equation}}
\setcounter{equation}{0}
\renewcommand{\thefigure}{S\arabic{figure}}
\setcounter{figure}{0}
\renewcommand{\thetable}{S\arabic{table}}
\setcounter{table}{0}

\section{The mean-field analysis}

\begin{figure}[b]
\centering
{
\includegraphics[width=7cm]{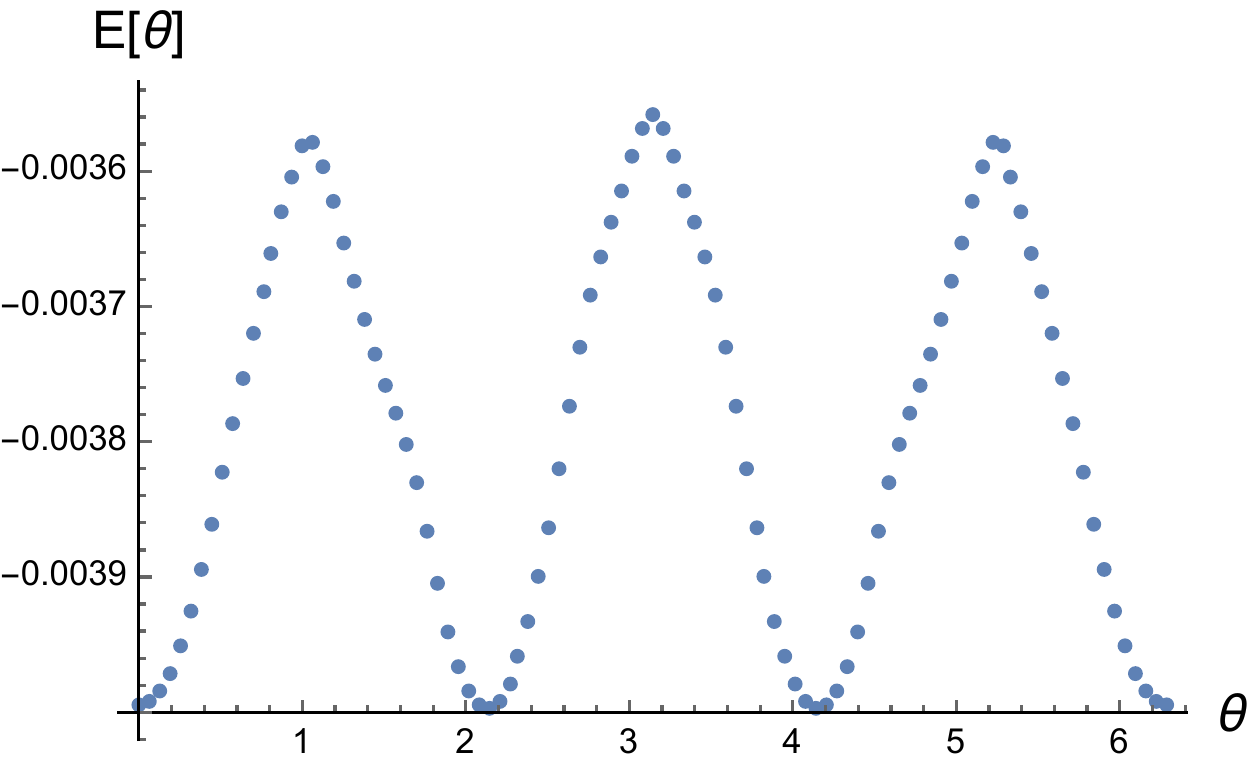}}
\caption{\label{E2order}The ground state energy in the ordered phase for $m=6$ and $U=7.8$ as a function of $\theta$, where the magnitude is fixed, $|\phi|=0.2$.}
\end{figure}

The order parameters for the nematic phase are given by $\phi_i(x)=\langle c^\dag(x) \sigma^i c(x) \rangle$. To explore the phase diagram, we set $t_1=t_2=t_3=1$ for $m=3.9$ and $m=4.1$. To check that the two-dimensional order parameter serves as an $E_2$ representation, $(\phi_1,\phi_2)=(|\phi| \cos\theta, |\phi| \sin\theta)$, we fix the magnitude of nematic order $\phi$ and plot the ground state energy as a function of $\theta$, as shown in Fig. \ref{E2order}. There are three degenerate ground states consistent with the transformation law of the $E_2$ representation, $\phi \rightarrow e^{i2\pi/3} \phi$.

We also plot the ground state energy as a function of order parameter across the transition. For the insulating system $m=3.9$, the transition from an insulator, in which two double-Weyl points were annihilated, to a nematic insulator is first-order. As shown in Fig. S\ref{insulator_nematic1}-S\ref{insulator_nematic3}, the ground-state energy as a function of the order parameter show a typical feature of presence of cubic terms in free energy, consistent with the Landau criterion. For a semi-metallic system $m=4.1$, the signature of the first-oder transition is strongly reduced by gapless fermions as shown in Fig. S\ref{dwf_nematic1}-S\ref{dwf_nematic3}. Note that the energy plotted in Fig. S\ref{dwf_nematic1}-S\ref{dwf_nematic3} is one order of magnitude smaller than that of Fig. S\ref{insulator_nematic1}-S\ref{insulator_nematic3}. Though it looks like a first-order transition, we show via RG calculations that it should be continuous.

\section{\label{non-analytical} Non-analytical terms of order parameter in double-Weyl semimetals}

Here we show the non-analytical terms arising by integrating out fermions explicitly. The zero-temperature free energy in the presence of a nonzero order parameter reads
\bea
	F[\phi]\propto-2N \int d^3 p \sqrt{(p_x^2-p_y^2 + \phi)^2+4p_x^2 p_y^2+ p_z^2}.
\eea
First we make a coordinate transformation, $p_x= \sqrt{q \sin\theta}\cos \varphi, p_y= \sqrt{q \sin\theta}\sin \varphi, p_z= q \cos \theta$, where $q$ is a positive variable. This transformation results in a nontrivial Jacobian, $|\frac{\partial p}{\partial q}|=\frac{q}{2}$. After that, we get
\bea
	F[\phi] &=& -\frac{1}{(2\pi)^3} \int_0^\pi d\theta \int_0^{2\pi} d\varphi \int_0^\pi dq~ \frac{q}{2} \sqrt{q^2+ 2q \phi \sin\theta \cos 2\varphi +\phi^2}.
\eea
The integration over $q$ is evaluated first, this results in a complicated integral. By expanding in the order of $\phi$, we get the non-analytical terms
\bea
F[\phi] = \frac{1}{(2\pi)^3}\int d\theta d\varphi \Big[\frac{5+3\cos 2\theta-6 \cos 4\varphi \sin^2 \theta}{48} |\phi|^3- \frac{\cos 2\varphi(3+\cos 2\theta-2 \cos 4\varphi \sin^2 \theta)}{16}\log(|\phi|+\cos 2\varphi \sin \theta \phi) \phi^3 \Big]. \nn \\
\eea
The integration can be evaluated directly, $F[\phi]=\frac{1}{18\pi^2} |\phi|^3 $.

\begin{figure}[t]
\centering
\subfigure[]{\label{insulator_nematic1}
\includegraphics[width=4.5cm]{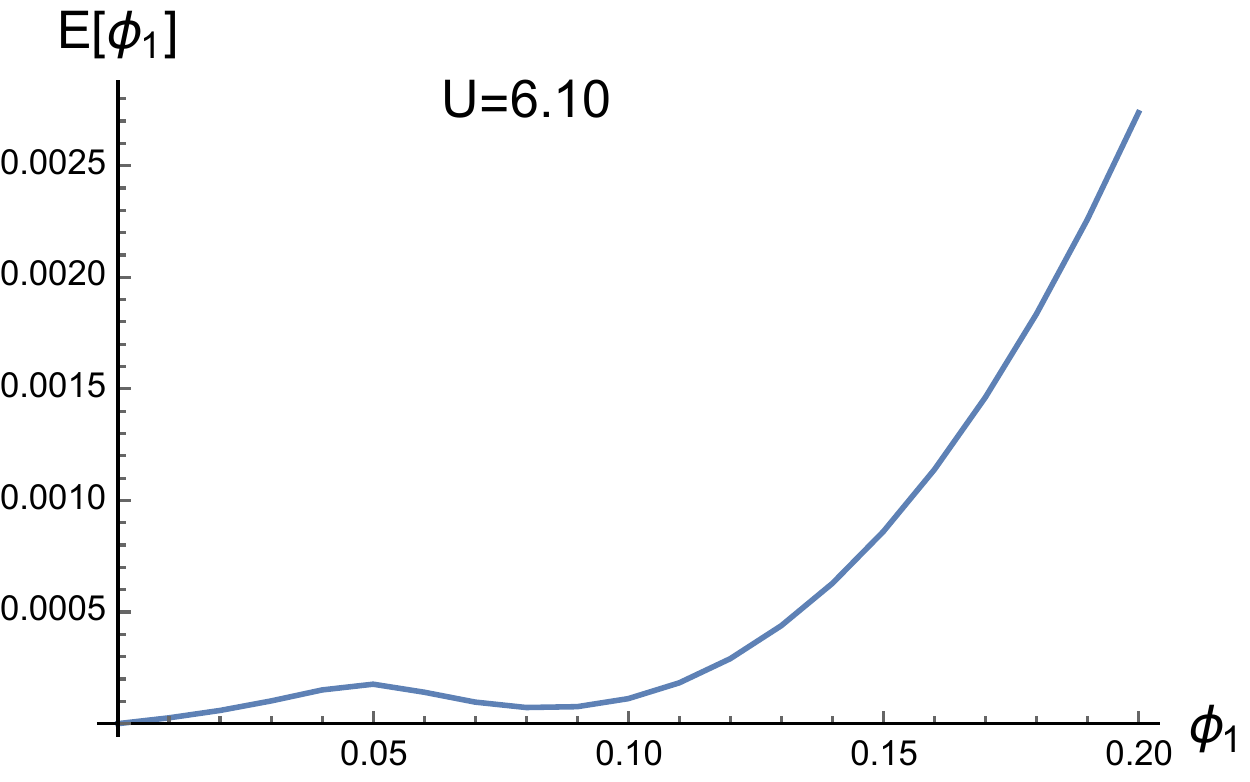}}
\subfigure[]{\label{insulator_nematic2}
\includegraphics[width=4.5cm]{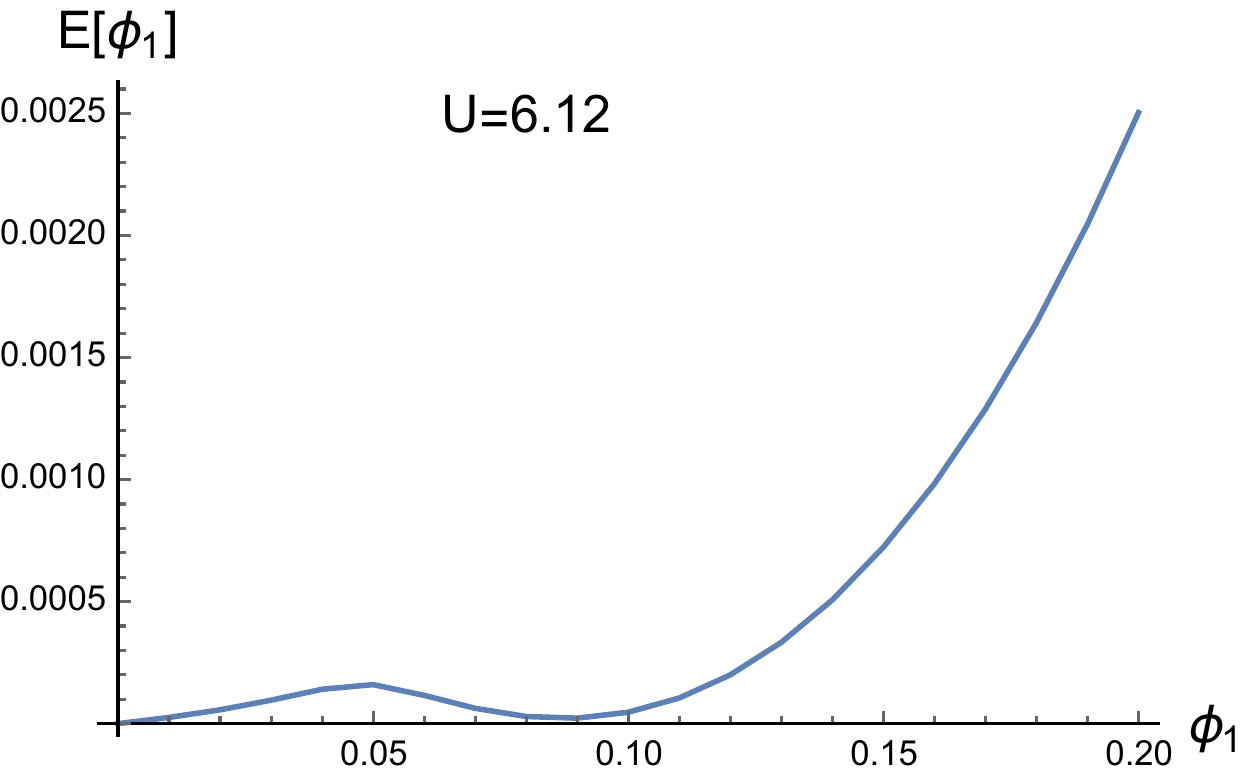}}
\subfigure[]{\label{insulator_nematic3}
\includegraphics[width=4.5cm]{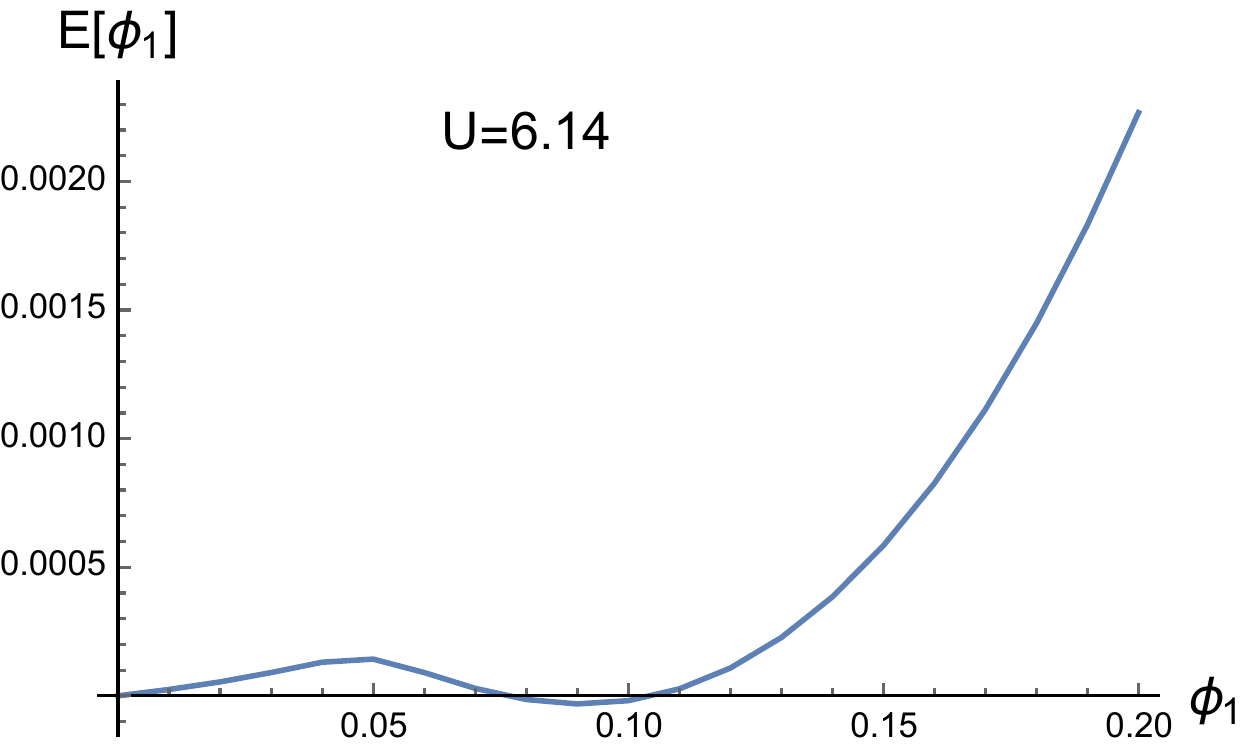}}
\subfigure[]{\label{dwf_nematic1}
\includegraphics[width=4.5cm]{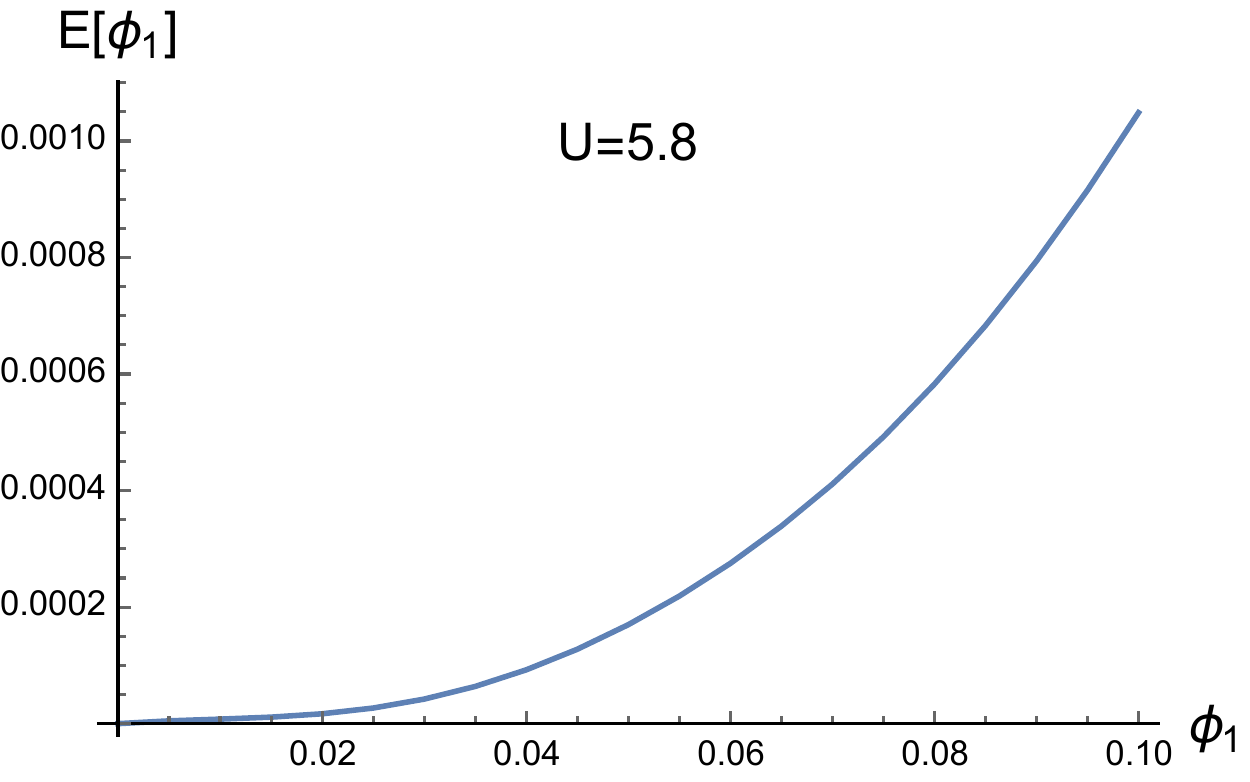}}
\subfigure[]{\label{dwf_nematic2}
\includegraphics[width=4.5cm]{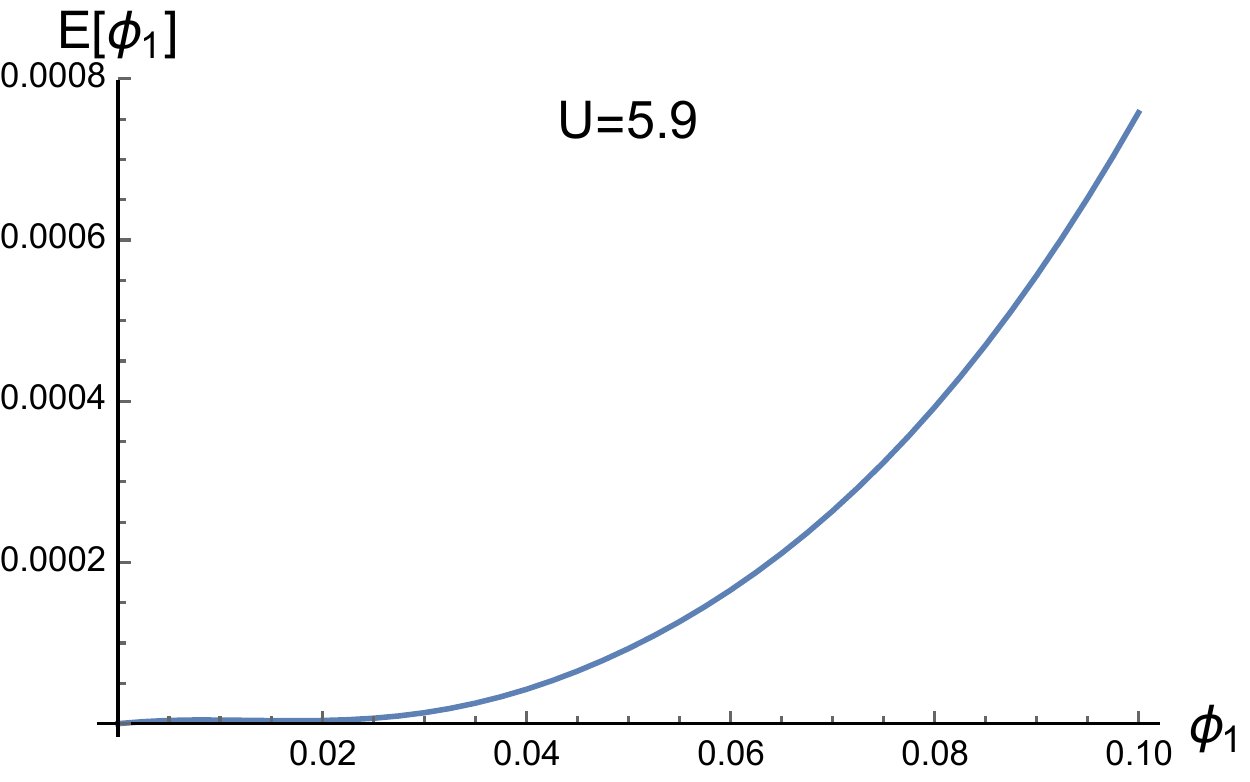}}
\subfigure[]{\label{dwf_nematic3}
\includegraphics[width=4.5cm]{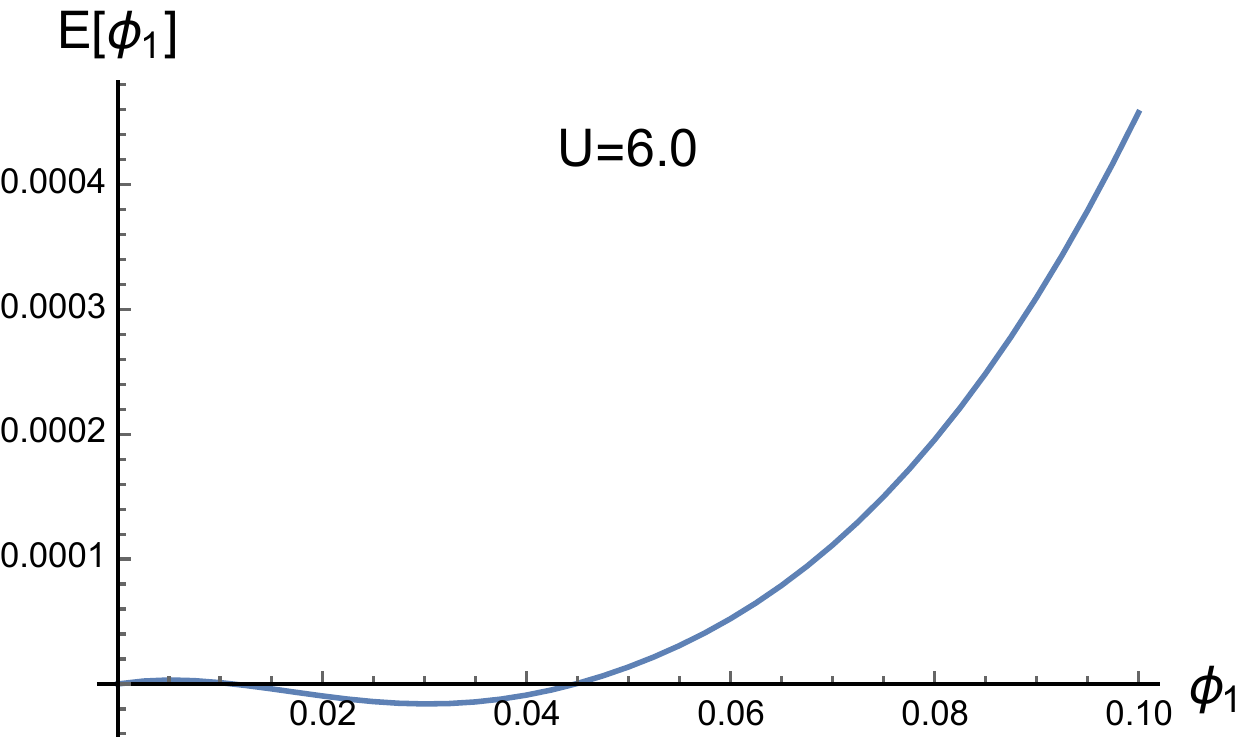}}
\caption{\label{dwf_nematic} Part (a)-(c) show the ground state energy as a function of order parameter in a transition from an insulator ($m=3.9$) to a nematic insulator. Part (d)-(f) show
the ground state energy as a function of order parameter across the transition from a double-Weyl semimetal ($m=4.1$) to a nematic Weyl semimetal.}
\end{figure}

\section{\label{rg_detail} Details for the renormalization equations at $Z_3$ nodal nematic transition of double-Weyl semimetals}

\begin{figure}[b]
	\subfigure[]{\label{kb1}
		\includegraphics[width=2.6cm]{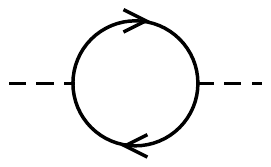}}
	\subfigure[]{\label{kb2}
		\includegraphics[width=2.7cm]{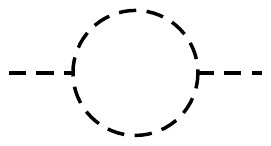}}
	\subfigure[]{\label{kf}
		\includegraphics[width=2.7cm]{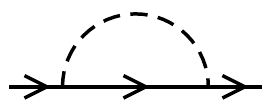}}
	\subfigure[]{\label{3b}
		\includegraphics[width=2.6cm]{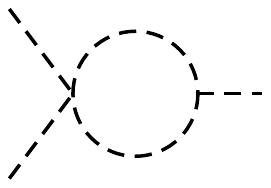}} \\
	\subfigure[]{\label{4b1}
		\includegraphics[width=2.6cm]{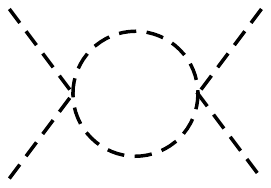}}
	\subfigure[]{\label{4b2}
		\includegraphics[width=2.5cm]{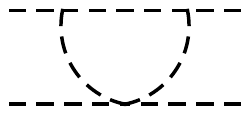}}
	\subfigure[]{\label{4b3}
		\includegraphics[width=2.2cm]{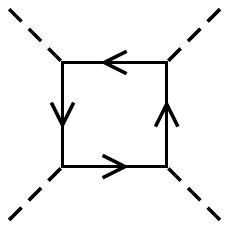}}
	\subfigure[]{\label{4b4}
		\includegraphics[width=2.2cm]{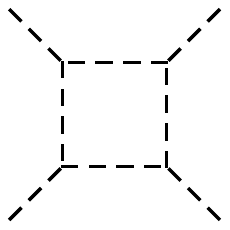}}
	\caption{\label{fig2} One-loop Feynman diagrams. The arrowed solid line indicates fermion propagator and dashed line indicates boson propagator.}
\end{figure}

The Feynman diagram Fig. S\ref{kf} gives fermion self-energy,
\bea
	\Sigma(p) = -\frac{1}{2} \!\times\! 2 g^2 \int_k (\Gamma^+ S(k)  \Gamma^- +\Gamma^- S(k)  \Gamma^+)D(p-k)
	= \frac{g^2l}{v_{b\perp}^2 v_{f3}} \Big[F_\omega(-i\omega_p) + F_z  v_{f3} p_z \Gamma^3 \Big],
\eea
where $\Gamma^\pm \!=\!\sigma^\pm$, $\Gamma^3\!=\!\sigma^z\tau^z$, $k_\perp\!=\! \sqrt{k_x^2\!+\!k_y^2}$, $\int_k \!\equiv\! \int_{-\infty}^\infty \frac{d\omega_k}{2\pi} \int \frac{d^3 k}{(2\pi)^3}$ and $S(k)$, $D(k)$ are fermion and boson propagators, respectively. Note that in the calculation, integrations of $\omega_k$ and $k_z$ are not constrained, while those of momentum $k_{x,y}$ are constrained in the momentum shell, i.e., $\Lambda_\perp e^{-l}\!<\! k_\perp \! <\!\Lambda_\perp$, where $\Lambda_\perp$ is a momentum cutoff in $k_x k_y$-plane and $l\!>\!0$ is the flow parameter. During the evaluation of Feynman diagrams, we have made a variable transformation, i.e., $k_{x,y}= \frac{v_{b\perp}}{A} q_{x,y}$, $k_z=\frac{v_{b\perp}^2}{Av_{f3}} q_z$ and $\omega_k= \frac{v_{b\perp}^2}{A} \omega_q$, and it is easy to check that $(\omega_q, \vec q)$ are dimensionless variables. $F_\omega$ and $F_z$ are given by
\bea
	F_\omega &=& \int_q \frac{2 \omega_q^2}{l(\omega_q^2+q_\perp^4+q_z^2)(\omega_q^2+q_\perp^2+\frac{v_{b3}^2}{v_{f3}^2}q_z^2)^2}=\frac{1-\alpha^2+ \alpha^2 \log \alpha^2}{8\pi^2 (1-\alpha^2)^2}+ O(\lambda), \\
	F_z &=& \frac{v_{b3}^2}{v_{f3}^2} \int_q \frac{2q_z^2}{l(\omega_q^2+q_\perp^4+q_z^2)(\omega_q^2+q_\perp^2+\frac{v_{b3}^2}{v_{f3}^2}q_z^2)^2} = \frac{1-\alpha^2+ \alpha^2 \log \alpha^2}{8\pi^2 (1-\alpha^2)^2}+ O(\lambda),
\eea
where $q_\perp=\sqrt{q_x^2+q_y^2} $ and $\int_q =\int_{-\infty}^\infty \frac{d \omega_q}{2\pi} \int_{-\infty}^\infty \frac{d q_z}{2\pi} \int_{\alpha e^{-l}}^\alpha \frac{d^2 q}{(2\pi)^2}$, and $\alpha= \frac{A\Lambda_\perp}{v_{b\perp}}$ is the cutoff in momentum $q_\perp$,
and $\lambda$ is a function of $\frac{v_{b3}}{v_{f3}}$ that will be defined below.

The boson self-energy is given by Feynman diagram Fig. S\ref{kb1} and S\ref{kb2}. Evaluation of Feynman diagram in Fig. S\ref{kb1} gives
\bea
	\Pi^{(1)}(p) \!&=&\! \frac{g^2}{2} \!\int_k\! \text{Tr} \Big[ \Gamma^+ S(k\!+\!p) \Gamma^- S(k)\!+\!\Gamma^- S(k\!+\!p) \Gamma^+ S(k) \Big] =  \frac{4N g^2 l}{v_{b\perp}^2 v_{f3}} \Big[G^{(1)}_\omega \omega_p^2  +G^{(1)}_\perp v_{b\perp}^2 p_\perp^2 +G^{(1)}_z v_{b3}^2 p_z^2  \Big],
\eea
where $\Tr$ is the trace in Gamma matrices and flavor space, and $\Tr 1 \!=\! 4N$, where we have also promoted the flavors of four-component double-Weyl fermions to be $N$. Evaluation of Feynman diagram Fig. S\ref{kb2} gives
\bea
	\Pi^{(2)}(p) = -\frac{1}{2} \times 36 b^2 \int_k D(k)D(k+p) =\frac{b^2 A^2 l}{v_{b\perp}^6v_{f3}} \Big[ G_\omega^{(2)}  \omega_p^2 \!+\! G_\perp^{(2)} v_{b\perp}^2 p_\perp^2 \!+\! G_z^{(2)} v_{b3}^2 p_z^2  \Big].
\eea
The the boson self-energy is  $\Pi(p)= \Pi^{(1)}(p)+ \Pi^{(2)}(p)$. Those $G^{(1)}_i$ are given by
\bea
	G_\omega^{(1)} &=& \frac{1}{4l} \!\int_q \!\Big[ \!\frac{6\omega_q^2+2q_z^2}{(\omega_q^2+ q_\perp^4+ q_z^2)^3} \!-\! \frac{8\omega_q^2(q_z^2+\omega_q^2)}{(\omega_q^2+ q_\perp^4+ q_z^2)^4}\Big] \!=\! \frac{1}{48\pi^2 \alpha^2}, \\
	G^{(1)}_\perp &=& \frac{1}{4l} \int_q \Big[ \frac{8q_\perp^2(\omega_q^2+q_z^2)}{(\omega_q^2+ q_\perp^4+ q_z^2)^3} \!-\! \frac{16 q_\perp^6(q_z^2+\omega_q^2)}{(\omega_q^2+ q_\perp^4+q_z^2)^4} \Big] \!=\! \frac{1}{24\pi^2}, \\
	G^{(1)}_z &=& \frac{1}{4l}\! \int_q\! \Big[ \!\frac{6q_z^2+2\omega_q^2}{(\omega_q^2+q_\perp^4+  q_z^2)^3}\! -\! \frac{8q_z^2(q_z^2+\omega_q^2)}{(\omega_q^2+q_\perp^4+ q_z^2)^4}\Big] \!=\! \frac{1}{48\pi^2 \alpha^2},
\eea
and $G^{(2)}_i$ are given by
\bea
	G^{(2)}_\omega &=& -\frac{9}{l} \int_q \Big[ \frac{8 \omega_q^2}{(\omega_q^2+q_\perp^2+ \frac{v_{b3}^2}{v_{f3}^2} q_z^2)^4}- \frac{2}{(\omega_q^2+q_\perp^2+ \frac{v_{b3}^2}{v_{f3}^2} q_z^2)^3}\Big] = \frac{3}{8\pi^2 \alpha^2}+ O(\lambda),\\
	G^{(2)}_\perp &=& -\frac{9}{l} \int_q \Big[ \frac{4 q_\perp^2}{(\omega_q^2+q_\perp^2+ \frac{v_{b3}^2}{v_{f3}^2} q_z^2)^4}- \frac{2}{(\omega_q^2+q_\perp^2+ \frac{v_{b3}^2}{v_{f3}^2} q_z^2)^3}\Big] = -\frac{3}{8\pi^2 \alpha^2}+ O(\lambda),\\
	G^{(2)}_z &=& -\frac{9}{l}\int_q \Big[ \frac{8 \frac{v_{b3}^2}{v_{f3}^2} q_z^2}{(\omega_q^2+q_\perp^2+ \frac{v_{b3}^2}{v_{f3}^2} q_z^2)^4}- \frac{2}{(\omega_q^2+q_\perp^2+ \frac{v_{b3}^2}{v_{f3}^2} q_z^2)^3}\Big]= \frac{3}{8\pi^2 \alpha^2}+ O(\lambda).
\eea
The full set of RG equations for the various constants appearing in the kinetic energy part are given by
\bea
	\frac{d\log A}{dl} &=& z-2z_1- \frac{g^2}{v_{b\perp}^2 v_{f3}}F_\omega,  \\
	\frac{d\log v_{b\perp}}{dl} &=& z-z_1+ \frac{\text{Tr} 1 g^2}{2v_{b\perp}^2 v_{f3}}(G_\perp^{(1)}- G_\omega^{(1)}) + \frac{b^2 A^2}{2v_{b\perp}^6v_{f3}}(G_\perp^{(2)}- G_\omega^{(2)}) , \\
	\frac{d\log v_{f3}}{dl} &=& z-1 +\frac{g^2}{v_{b\perp}^2 v_{f3}}(F_z-F_\omega), \label{vf3}\\
	\frac{d\log v_{b3}}{dl} &=& z-1 + \frac{\text{Tr} 1 g^2}{2v_{b\perp}^2 v_{f3}}(G_z^{(1)}- G_\omega^{(1)}) + \frac{b^2 A^2}{2v_{b\perp}^6v_{f3}}(G_z^{(2)}- G_\omega^{(2)}) \label{vb3}.
\eea
From above RG equations, we have
\bea
	\frac{d\log (v_{b3}/v_{f3})}{dl} &=& -\frac{g^2}{v_{b\perp}^2 v_{f3}}(F_z-F_\omega)+ \frac{b^2 A^2}{2v_{b\perp}^6v_{f3}}(G_z^{(2)}- G_\omega^{(2)}).
\eea
Setting $\frac{v_{b3}}{v_{f3}}\!=\!1\!+\!\lambda$, and assuming $\lambda \!\ll\! 1$, a simple manipulation leads to $\frac{d \lambda}{dl}\!=\! -\Delta_\lambda \lambda$, where $\Delta_\lambda\equiv \frac{g^2}{2v_{b\perp}^2 v_{f3}}H_1\!+\! \frac{b^2 A^2}{2v_{b\perp}^6v_{f3}} H_2$ with
\bea
	H_1&=& \frac{1}{l} \int_q \frac{4q_z^2( q_\perp^2+3\omega_q^2-q_z^2)}{(\omega_q^2+q_z^2+ q_\perp^2)^3(\omega_q^2+q_z^2+q_\perp^4)}=\frac{1}{l}\int \frac{dq_x dq_y}{(2\pi)^2} \int_0^\infty \frac{q_0 d q_0}{2\pi} \frac{2q_0^2( q_\perp^2+q_0^2)}{(q_0^2+ q_\perp^2)^3(q_0^2+q_\perp^4)}, \\
	H_2 &=& \frac{1}{l} \int_q \frac{144q_z^2( q_\perp^2+5 \omega_q^2 -3 q_z^2)}{(\omega_q^2+q_z^2+ q_\perp^2)^5} = \frac{1}{l}\int \frac{dq_x dq_y}{(2\pi)^2} \int_0^\infty \frac{q_0 d q_0}{2\pi} \frac{72q_0^2( q_\perp^2+ q_0^2)}{(q_0^2+ q_\perp^2)^5},
\eea
where we use the rotational symmetry between $\omega_q$ and $q_z$ in the above integration to deduce that both $H_1$ and $H_2$ are positive. As a consequence, $\lambda=0$ is a stable fixed point. The RG equation for boson velocity in $xy$-plane $v_{b\perp}$ reads
\bea
	\frac{d\log v_{b\perp}}{dl} &=& 1-z_1-\frac38 \beta+N (2\alpha^2-1) \gamma, \label{vbp}
\eea
where $\alpha\!=\!\frac{A\Lambda_\perp}{v_{b\perp}}$ is also a dimensionless constant. In order to maintain the velocity, one gets $z_1\!=\!1\!-\!\frac{3}{8}\beta+N (2\alpha^2-1) \gamma$ at the fixed point.

Next, we calculate the remaining Feynman diagrams corresponding to coupling constant renormalizations. The Feynman diagram in Fig. S\ref{3b} gives
\bea
	\Gamma_{\phi^3}=\Gamma_{\phi^{*3}}= -\frac{3l}{4\pi^2}\frac{b u }{v_{b\perp}^2v}.
\eea
The Feynman diagrams in Fig. S\ref{4b1}, S\ref{4b2}, S\ref{4b3} and S\ref{4b4} give
\bea
	\Gamma_{|\phi|^4} &=& -\frac{5l}{4\pi^2} \frac{u^2}{v_{b\perp}^2 v}+ \frac{9l}{\pi^2 } \frac{u b^2}{v_{b\perp}^4 v \Lambda^2_\perp} +\frac{N l}{96\pi^2}\frac{g^4}{A^2 v \Lambda_{\perp}^2}-\frac{27l}{2\pi^2} \frac{b^4}{v_{b\perp}^{6} v\Lambda^4_\perp}.
\eea
Introducing the dimensionless coupling constants, namely, $\beta=\frac{b^2}{\pi^2 v_{b\perp}^4 v \Lambda_\perp^2}$, $\gamma =\frac{g^2}{24\pi^2 A^2 v\Lambda_\perp^2}$ and $\delta= \frac{u}{\pi^2 v_{b\perp}^2 v}$, the RG equations read
\bea
	\frac{d \alpha}{dl} &=& (-1+\frac34 \beta+2 N \gamma) \alpha- 4N \gamma \alpha^3 + \frac{3\gamma(\alpha^2-1-\alpha^2 \log \alpha^2)\alpha^3}{(\alpha^2-1)^2}, \\
	\frac{d \beta}{dl} &=& (2-4N\gamma-\frac32 \delta)\beta-\frac38 \beta^2-4N \alpha^2 \gamma, \\
	\frac{d \gamma}{dl} &=& 2\gamma-4N\gamma-\frac98 \beta\gamma + 4N \alpha^2 \gamma^2, \\
	\frac{d \delta}{dl} &=& (9\beta-2N\gamma) \delta- \frac54 \delta^2 - \frac{27}{2} \beta^2 -4N \alpha^2 \gamma \delta+ 6N\alpha^2 \gamma^2.
\eea
This RG equations can be solved by a stable fixed point $(\alpha^*,\beta^*,\gamma^*,\delta^*)=(0,0,\frac{1}{2N},0)$. By expanding the RG equations near $\alpha=0$, one gets
\bea
	\frac{d \alpha}{dl} =-(2+ \frac{3}{2N}) \alpha^3.
\eea
Above RG equation shows that $\alpha=0$ is marginally stable at this fixed point, and as a result, we expand the RG equations in the order of $\alpha$, as shown in the main text.

\end{widetext}

\end{appendix}

\end{document}